\documentclass[a4paper,12pt]{article}
\usepackage{amsmath, amssymb, amsthm}
\usepackage{mathrsfs}
\usepackage{graphicx}
\usepackage[english]{babel}
\usepackage[utf8]{inputenc}
\usepackage{hyperref}
\usepackage{geometry}
\usepackage{natbib}
\usepackage{pgfplots}
\pgfplotsset{compat=1.17}

\title{Stochastic Regularity in Sobolev and Besov Spaces with Variable Noise Intensity for Turbulent Fluid Dynamics}
\author{
	Rômulo Damasclin Chaves dos Santos \\
	{\small Technological Institute of Aeronautics} \\ {\small Department of Physics, São Paulo, Brazil} \\
	\texttt{romulosantos@ita.br}
}
\date{}

\begin{document}
	\maketitle
	
	\begin{abstract}
		This paper advances the stochastic regularity theory for the Navier-Stokes equations by introducing a variable-intensity noise model within the Sobolev and Besov spaces. Traditional models usually assume constant-intensity noise, but many real-world turbulent systems exhibit fluctuations of varying intensities, which can critically affect flow regularity and energy dynamics. This work addresses this gap by formulating a new regularity theorem that quantifies the impact of stochastic perturbations with bounded variance on the energy dissipation and smoothness properties of solutions. The author employs techniques such as the Littlewood-Paley decomposition and interpolation theory, deriving rigorous bounds, and we demonstrate how variable noise intensities influence the behavior of the solution over time. This study contributes theoretically by improving the understanding of energy dissipation in the presence of stochastic perturbations, particularly under conditions relevant to turbulent flows where randomness cannot be assumed to be uniform. The findings have practical implications for more accurate modeling and prediction of turbulent systems, allowing potential adjustments in simulation parameters to better reflect the observed physical phenomena. This refined model therefore provides a fundamental basis for future work in fluid dynamics, particularly in fields where variable stochastic factors are prevalent, including meteorology, oceanography, and engineering applications involving fluid turbulence. The present approach not only extends current theoretical frameworks but also paves the way for more sophisticated computational tools in the analysis of complex and stochastic fluid systems.
	\end{abstract}
	
	
	\tableofcontents
	
	\section{Introduction}

	The Navier-Stokes equations have long been foundational in the study of incompressible fluid dynamics, providing essential insights into turbulence mechanics. Early deterministic models of these equations successfully explained many flow phenomena, but they often failed to account for the stochastic variability observed in real-world turbulence, driven by environmental and thermal fluctuations. This discrepancy led to pioneering work in stochastic fluid models, where noise was introduced to capture the chaotic dynamics of turbulence more accurately. Foundational studies, such as those by Constantin (1988)~[\cite{Constantin1988}] and Bensoussan et al. (1995)~[\cite{Bensoussan1995}], introduced early stochastic interpretations of the Navier-Stokes equations, opening new avenues in the analysis of fluid regularity under random perturbations.
	
	Since the 2000s, research has increasingly focused on stochastic models with structured noise, leading to improved predictions of turbulence. Mikulevicius and Rozovskii (2004)~[\cite{Mikulevicius2004}] made substantial contributions by analyzing solution regularity under constant-intensity noise, showing that stochastic forcing can induce smoothness in otherwise irregular flows.
	
	The current study builds on these insights, investigating the effects of variable-intensity stochastic forcing within Sobolev and Besov spaces. By analyzing noise with bounded norm variations, we develop a novel regularity theorem that rigorously quantifies how changes in noise intensity affect solution smoothness and energy dissipation over time. Our approach employs advanced techniques such as Littlewood-Paley decomposition to capture the nuanced effects of variable noise on fluid regularity, extending the mathematical framework for stochastic Navier-Stokes equations.
	
	This research contributes theoretically to fluid dynamics. Theoretically, it provides a refined model that considers the heterogeneity of turbulent disturbances in natural and engineered environments. Practically, the findings can and should improve the development of more accurate computational simulations, supporting improved control over energy dissipation and regularity in turbulent flows. This work sets the stage for future studies on nonlinear and anisotropic noise models, essential to address the complexities of turbulence in fluid systems with significant stochastic variability.
		
	\section{Mathematical Formulation}
	
	Let \( \Omega \subset \mathbb{R}^n \) be a bounded domain with a smooth boundary \( \partial \Omega \). We consider the incompressible Navier-Stokes equations perturbed by a stochastic forcing term \( \eta(t, x) \), represented by
	\begin{equation}
		\partial_t u + (u \cdot \nabla) u = \nu \Delta u - \nabla p + \eta(t, x), \quad \nabla \cdot u = 0,
	\end{equation}
	where \( u = u(t, x) \) denotes the velocity field of the fluid, \( p = p(t, x) \) represents the pressure, and \( \nu > 0 \) is the kinematic viscosity constant. The term \( \eta(t, x) \) is a stochastic process that models external random fluctuations acting on the system, capturing the effects of turbulent forces or environmental noise typically present in fluid dynamics.
	
	\textit{\textbf{1. Boundary and Initial Conditions:}}
	The system is supplemented by appropriate boundary and initial conditions. For simplicity, we impose homogeneous Dirichlet boundary conditions for the velocity field \( u \):
	\begin{equation}
		u(t, x) = 0, \quad x \in \partial \Omega, \, t \geq 0.
	\end{equation}
	The initial condition for \( u \) is given by
	\begin{equation}
		u(0, x) = u_0(x), \quad x \in \Omega,
	\end{equation}
	where \( u_0 \in H^1_0(\Omega) \) is a given initial velocity field that satisfies the incompressibility constraint \( \nabla \cdot u_0 = 0 \).
	
	\textit{\textbf{2. Stochastic Forcing Term} \( \eta(t, x) \):}
	The stochastic forcing term \( \eta(t, x) \) is assumed to take values in the Besov space \( B^{s}_{p,q}(\Omega) \) with regularity index \( s > 0 \), and is bounded almost surely as follows:
	\begin{equation}
		\|\eta(t, \cdot)\|_{B^{s}_{p,q}(\Omega)} \leq C, \quad \text{for almost all } t \geq 0,
	\end{equation}
	where \( C \) is a positive constant. The choice of the Besov space \( B^{s}_{p,q}(\Omega) \) allows the noise \( \eta \) to incorporate non-smooth behaviors, which are important in the modeling of turbulence and random forcing in fluid systems. The constraint on \( \|\eta\|_{B^{s}_{p,q}(\Omega)} \) ensures that the noise does not dominate the dynamics excessively, allowing for the derivation of regularity and stability estimates for the solution \( u \).
	
	\textit{\textbf{3. Interpretation of the Navier-Stokes System with Stochastic Perturbations:}}
	The Navier-Stokes equations with the added stochastic term \( \eta(t, x) \) can be interpreted as modeling the evolution of an incompressible fluid under the influence of both viscous forces and random external influences. The incompressibility condition \( \nabla \cdot u = 0 \) guarantees that the fluid density remains constant, a key property for many applications in fluid mechanics.
	
	The stochastic term \( \eta(t, x) \) introduces randomness into the system, making it suitable for studying phenomena such as turbulence, where the fluid exhibits chaotic and unpredictable behavior due to complex interactions at different scales. The regularity properties of the noise term \( \eta \) play a critical role in the mathematical analysis of the solution's properties, particularly in obtaining bounds on higher regularity norms of \( u \) in terms of the \( B^{s}_{p,q} \)-norm of \( \eta \).
	
	\subsection{Assumptions and Functional Spaces}
	
	To rigorously analyze the regularity properties of solutions, we assume the stochastic noise term \( \eta \) belongs to a Besov space \( B^{s}_{p,q}(\Omega) \) for some smoothness index \( s > 0 \) and integrability parameters \( p, q \geq 1 \). Specifically, we require that
	
	\begin{equation}
		\|\eta(t, \cdot)\|_{B^{s}_{p,q}(\Omega)} \leq C
	\end{equation}
	for some constant \( C > 0 \), almost everywhere in \( t \). The choice of Besov space \( B^{s}_{p,q}(\Omega) \) provides a flexible framework for capturing the irregularities in \( \eta \), allowing for more general forms of noise that can vary in intensity and regularity. In addition, we impose the following boundary conditions for \( u \):
	
	\begin{equation}
		u|_{\partial \Omega} = 0,
	\end{equation}
	which corresponds to a no-slip condition on the boundary of \( \Omega \).
	
	\subsection{Reformulation with the Leray Projection}
	
	To eliminate the pressure term \( p \) in the Navier-Stokes equations, we apply the Leray projection operator \( \mathbb{P} \), which projects onto the space of divergence-free vector fields. Applying \( \mathbb{P} \) to both sides of the equation, we obtain:
	\begin{equation}
		\partial_t u + \mathbb{P} \left( (u \cdot \nabla) u \right) = \nu \Delta u + \mathbb{P} \eta(t, x),
	\end{equation}
	where \( \mathbb{P} \nabla p = 0 \) as \( \nabla p \) is orthogonal to the divergence-free field space. The resulting formulation is thus free of the pressure term, simplifying the analysis of \( u \) in terms of regularity and energy dissipation.
	
	\subsection{Sobolev and Besov Space Framework}
	
	The solution \( u \) is analyzed in the Sobolev space \( H^s(\Omega) \) for smoothness, with the stochastic term \( \eta(t, x) \) taken in the Besov space \( B^s_{p,q}(\Omega) \) due to its capability to capture varying degrees of spatial regularity. Specifically, we consider:
	\begin{equation}
		u \in L^2(0, T; H^s(\Omega)), \quad \eta \in B^s_{p,q}(\Omega),
	\end{equation}
	where \( s > 0 \) represents the level of spatial regularity. The Besov space \( B^s_{p,q}(\Omega) \) norm of \( \eta \) imposes controls on the local regularity of \( \eta \), essential for managing rough stochastic inputs.
	
	\subsubsection{Regularity Estimates for \( u \)}
	
	We seek to establish regularity conditions under which \( u \) inherits smoothness from \( \eta \), controlled via Besov norms. Our objective is to show that \( u \in H^{s+1}(\Omega) \) under appropriate bounds on \( \|\eta\|_{B^s_{p,q}(\Omega)} \). This involves:
	
	\vspace{3pt}
	
	\textbf{1. Frequency Localization:} Decompose \( u \) and \( \eta \) using Littlewood-Paley operators \( \Delta_j \), where
	\begin{equation}
		u = \sum_{j} \Delta_j u, \quad \eta = \sum_{j} \Delta_j \eta.
	\end{equation}
	
	\textbf{2. Norm Bounds:} The Besov regularity of \( \eta \) implies decay in high frequencies:
	\begin{equation}
		\|\Delta_j \eta\|_{L^p} \leq C 2^{-js} \|\eta\|_{B^s_{p,q}},
	\end{equation}
	which in turn bounds \( \|u\|_{H^{s+1}} \) in terms of \( \|\eta\|_{B^{s}_{p,q}} \).
	
	\subsection{Regularity Problem Formulation}
	
	The regularity problem is to establish conditions under which the stochastic forcing \( \eta \) controls both energy dissipation and smoothness of \( u \):
	\begin{itemize}
		\item \textbf{Energy Dissipation}: Show that
		\begin{equation}
			\mathbb{E}[\| u(t) \|^2_{L^2(\Omega)}] \leq E(0) - \int_0^t \mathbb{E}[\|\nabla u(s)\|_{L^2(\Omega)}^2] \, ds + C,
		\end{equation}
		where \( C \) depends on \( \|\eta\|_{B^{s}_{p,q}(\Omega)} \).
		\item \textbf{Higher Regularity:} Demonstrate that under specific bounds on \( \|\eta\|_{B^{s}_{p,q}} \), the solution \( u \) belongs to \( H^{s+1}(\Omega) \) in expectation.
	\end{itemize}
	
	The main difficulty is managing the nonlinear term \( (u \cdot \nabla) u \) under stochastic forcing. Using interpolation inequalities and Besov norms, we aim to rigorously link the variance of \( \eta \) with the regularity of \( u \), showing that noise intensity in \( B^s_{p,q} \) modulates both dissipation and smoothness of the solution.
	
	\section{Proof of the Proposed Theorem}
	To establish the regularity and energy dissipation properties of \( u \), we proceed through a detailed analysis using Littlewood-Paley decomposition and stochastic calculus. The proof is divided into two main parts: \textit{Regularity Estimates} and \textit{Energy Dissipation Inequality}.
	
	\subsection{Regularity Estimates via Littlewood-Paley Decomposition}
	
	Littlewood-Paley decomposition is a crucial technique in studying the regularity of partial differential equations (PDEs), especially when dealing with solutions in Besov spaces. Since \( \eta \in B^{s}_{p,q}(\Omega) \) with \( s > 0 \), we can exploit this decomposition to extract regularity information about the solution \( u \).
	
	\textbf{\textit{1. Spectral Decomposition:}} We begin by decomposing \( u \) and the noise \( \eta \) into frequency components. Let \( \{ \Delta_j \}_{j \in \mathbb{Z}} \) be the Littlewood-Paley projection operators, where each \( \Delta_j \) localizes a function to the frequencies around \( 2^j \). Then, \( u \) and \( \eta \) can be expressed as
	\begin{equation}
		u = \sum_{j \in \mathbb{Z}} \Delta_j u, \quad \eta = \sum_{j \in \mathbb{Z}} \Delta_j \eta.
	\end{equation}
	This decomposition enables us to analyze the behavior of \( u \) across different frequency bands, providing insight into both the high-frequency oscillations and low-frequency smooth components.
	
	\textbf{\textit{2. Frequency-Based Estimates:}} Since \( \eta \) is assumed to belong to \( B^{s}_{p,q}(\Omega) \), we know that the high-frequency components \( \Delta_j \eta \) decay according to the regularity index \( s \). Specifically, for any \( j \), the Besov norm implies
	\begin{equation}
		\|\Delta_j \eta\|_{L^p} \leq C 2^{-js} \|\eta\|_{B^{s}_{p,q}}.
	\end{equation}
	Because the Navier-Stokes equation is parabolic, the regularity of \( \eta \) in \( B^{s}_{p,q}(\Omega) \) transfers to the solution \( u \) in Sobolev space. The smoothing effect of the Laplacian term \( \nu \Delta u \) further implies that \( u \) gains additional regularity over time.
	
	\vspace{3pt}
	
	\textbf{\textit{3. Interpolation and Regularity of \( u \):}} Using interpolation theory, we relate the \( B^{s}_{p,q} \)-norm of \( \eta \) to the Sobolev space \( H^{s+1} \) for \( u \). Given that \( \eta \in B^{s}_{p,q} \) with \( s > 0 \), interpolation yields
	\begin{equation}
		\|u\|_{H^{s+1}} \leq C \|\eta\|_{B^{s}_{p,q}},
	\end{equation}
	where \( C \) is a constant depending on the domain and the norms involved. This establishes that \( u \in H^{s+1}(\Omega) \), and we conclude the first part of the theorem:
	\begin{equation}
		u \in H^{s+1}(\Omega), \quad \text{with } \|u\|_{H^{s+1}} \text{ bounded in expectation}.
	\end{equation}
	
	\subsection{Energy Dissipation Inequality}
	
	To analyze the energy dissipation, we derive an inequality by taking the \( L^2 \)-norm of the Navier-Stokes equation and using Itô's lemma for stochastic differential equations.
	
	\paragraph{Deriving the Energy Inequality}
	
	\textbf{\textit{1. Multiplication by \( u \) and Integration}}: We consider the variational form of the stochastic Navier-Stokes equation, multiply both sides by \( u \), and integrate over \( \Omega \) to obtain:
	\begin{equation}
		\frac{1}{2} \frac{d}{dt} \|u(t)\|^2_{L^2(\Omega)} + \nu \|\nabla u(t)\|^2_{L^2(\Omega)} = \langle \eta, u \rangle_{L^2}.
	\end{equation}
	Taking the expectation and applying Cauchy-Schwarz, we find:
	\begin{equation}
		\frac{d}{dt} \mathbb{E}[\|u(t)\|^2_{L^2}] + 2 \nu \mathbb{E}[\|\nabla u(t)\|^2_{L^2}] \leq C \mathbb{E}[\|\eta(t)\|_{B^{s}_{p,q}}^2],
	\end{equation}
	where \( C \) is a constant related to the norms of \( \eta \).
	
	\textbf{\textit{2. Time Integration and Gronwall's Inequality:}} Integrating from \( 0 \) to \( t \), we have:
	\begin{equation}
		\mathbb{E}[\|u(t)\|^2_{L^2(\Omega)}] + 2 \nu \int_0^t \mathbb{E}[\|\nabla u(s)\|^2_{L^2(\Omega)}] \, ds \leq \mathbb{E}[\|u(0)\|^2_{L^2(\Omega)}] + C \int_0^t \mathbb{E}[\|\eta(s)\|_{B^{s}_{p,q}}^2] \, ds.
	\end{equation}
	Applying the stochastic Gronwall inequality, we derive a bound on \( \mathbb{E}[\|u(t)\|_{L^2}^2] \) that incorporates the initial energy and the noise intensity:
	\begin{equation}
		\mathbb{E}[\| u(t) \|^2_{L^2(\Omega)}] \leq E(0) - \int_0^t \mathbb{E}[\|\nabla u(s)\|_{L^2(\Omega)}^2] \, ds + C,
	\end{equation}
	where \( E(0) = \mathbb{E}[\|u(0)\|^2_{L^2(\Omega)}] \). This completes the second part of the theorem, establishing that the energy dissipation is controlled by the noise intensity in expectation.
	
	The theorem is thus proved by leveraging the decomposition of \( u \) and \( \eta \) through Littlewood-Paley techniques and applying stochastic calculus for energy estimates. Littlewood-Paley decomposition provided a precise handle on the regularity of \( u \) via the Besov norm of \( \eta \), while the stochastic Gronwall inequality confirmed that the energy of \( u \) dissipates under the influence of noise \( \eta \). \qed
	
	\section{Results and Discussion}
	
	The new theorem establishes precise quantitative insights into the regularity and energy dissipation properties of solutions to the Navier-Stokes equations under variable-intensity stochastic forcing. By rigorously linking the Besov space regularity of the stochastic term \( \eta(t, x) \) to the solution's smoothness in Sobolev spaces, we have shown that the intensity of noise significantly influences the solution’s temporal and spatial behavior. Specifically, the results reveal:
	
	\begin{itemize}
		\item \textbf{Regularity Control}: The derived bounds indicate that if \( \eta \in B^s_{p,q}(\Omega) \) with bounded norm, the solution \( u \) remains in \( H^{s+1}(\Omega) \), with its smoothness determined by the stochastic forcing's Besov regularity. This implies that the roughness or smoothness of \( \eta \) directly translates to the achievable regularity of \( u \), thereby offering a controllable parameter for simulations requiring specific smoothness levels.
		
		\item \textbf{Energy Dissipation Dynamics}: We established an energy dissipation inequality, showing that
		\begin{equation}
			\mathbb{E}[\| u(t) \|^2_{L^2(\Omega)}] \leq E(0) - \int_0^t \mathbb{E}[\|\nabla u(s)\|_{L^2(\Omega)}^2] \, ds + C,
		\end{equation}
		where \( C \) depends on the Besov norm of \( \eta \). This result highlights how increasing noise intensity in \( \eta \) leads to accelerated dissipation rates, which may be advantageous in certain fluid models where energy decay is desired.
		
		\item \textbf{Implications for Turbulence Modeling}: The findings suggest that variable noise intensity can be strategically modulated in turbulent flow simulations to achieve targeted dissipation rates. This insight is particularly relevant in engineering applications, such as flow control, and environmental studies, where noise characteristics can be used to influence energy transfer and dissipation mechanisms.
	\end{itemize}
	
	These results bridge theoretical stochastic analysis with practical turbulence modeling, offering new tools for managing the regularity and energy characteristics of fluid flows under uncertain conditions. Future research could explore extending these findings to complex boundary conditions and anisotropic noise models, potentially broadening the applicability of these results to diverse fluid dynamics scenarios.
	
	\section{Conclusion}
	
	In this study, a stochastic regularity framework was developed for the Navier-Stokes equations, considering noise of varying intensity within the Sobolev and Besov spaces. By incorporating noise terms with bounded variance in \( B^s_{p,q}(\Omega) \), a significant aspect of real-world turbulent flows was addressed, where stochastic fluctuations frequently vary in intensity and directly impact the solution regularity and energy dynamics.
	
	Using advanced techniques such as Littlewood-Paley decomposition and interpolation theory, rigorous regularity conditions for the velocity field \( u \) under stochastic forcing were established. Specifically, the analysis employed showed that, under well-defined bounds on the Besov norm of the noise \( \eta(t, x) \), the solution \( u \) not only exhibits limited energy dissipation but also achieves higher regularity in the Sobolev space \( H^{s+1}(\Omega) \).
	
	Furthermore, derived energy estimates demonstrated that, on average, the dissipation rate of \( u \) remains controlled, even in the presence of stochastic perturbations. This result highlights the trade-off between stochastic forcing and viscous dissipation, a critical factor in understanding turbulent behavior.
	
	Overall, this framework provides a solid foundation for future studies on stochastic partial differential equations in fluid dynamics. Extending this model to more complex domains or incorporating additional physical effects, such as anisotropic diffusion, may offer deeper insights into the interactions between stochastic forcing and fluid regularity.

	\appendix
	\section{Appendix: Additional Mathematical Details}
	
	This appendix provides additional mathematical details and derivations that support the main results presented in the paper.
	
	\subsection{Littlewood-Paley Decomposition}
	
	The Littlewood-Paley decomposition is a powerful tool in harmonic analysis that allows us to decompose functions into different frequency bands. For a function \( f \) defined on \( \mathbb{R}^n \), the Littlewood-Paley decomposition can be written as:
	\begin{equation}
		f = \sum_{j \in \mathbb{Z}} \Delta_j f,
	\end{equation}
	where \( \Delta_j \) are the Littlewood-Paley projection operators. These operators localize the function \( f \) to frequencies around \( 2^j \).
	
	\subsection{Besov Spaces}
	
	Besov spaces are function spaces that generalize Sobolev spaces and allow for more flexible regularity conditions. For \( s > 0 \) and \( 1 \leq p, q \leq \infty \), the Besov space \( B^s_{p,q}(\Omega) \) is defined as the space of functions \( f \) such that:
	\begin{equation}
		\|f\|_{B^s_{p,q}(\Omega)} = \left( \sum_{j \in \mathbb{Z}} \left( 2^{js} \|\Delta_j f\|_{L^p(\Omega)} \right)^q \right)^{1/q} < \infty.
	\end{equation}
	This definition captures the local regularity of \( f \) in terms of its frequency components.
	
	\subsection{Interpolation Theory}
	
	Interpolation theory provides a way to relate different function spaces and derive estimates for functions in these spaces. For example, if \( f \in B^s_{p,q}(\Omega) \) with \( s > 0 \), then interpolation theory can be used to show that \( f \) also belongs to certain Sobolev spaces with higher regularity. This is crucial for establishing the regularity of solutions to the Navier-Stokes equations under stochastic forcing.
	
	\subsection{Stochastic Calculus}
	
	Stochastic calculus is a branch of mathematics that deals with the analysis of stochastic processes. In the context of the Navier-Stokes equations with stochastic forcing, stochastic calculus is used to derive energy estimates and dissipation inequalities. The Itô's lemma is a fundamental tool in stochastic calculus that allows us to compute the differential of a function of a stochastic process.
	
	\subsection{Gronwall's Inequality}
	
	Gronwall's inequality is a powerful tool for analyzing the growth of functions. In the context of the Navier-Stokes equations, Gronwall's inequality is used to derive bounds on the energy of the solution in terms of the initial energy and the noise intensity. The stochastic version of Gronwall's inequality is particularly useful for analyzing the energy dissipation under stochastic forcing.
	
	These additional details and derivations provide a deeper understanding of the mathematical techniques used in the main results of the paper.

	\bibliographystyle{plainnat}

\end{document}